\documentclass[aps,prd,twocolumn,floatfix,noshowpacs,tightenlines,noshowkeys,superscriptaddress,amsmath,amssymb,nofootinbib]{revtex4}
\usepackage{amssymb,amsbsy,epsfig,color,graphicx}
\usepackage{color}
\usepackage{[longtable}
\usepackage{array}
\usepackage{dcolumn}   
\usepackage{cellspace}
\usepackage{mathtools}
\usepackage{amstext}
\usepackage{amssymb}
\usepackage{stmaryrd}
\usepackage{stackrel}
\usepackage{graphicx}
\usepackage{esint}
\usepackage[utf8]{inputenc}
\usepackage{blindtext}
\usepackage{float}
\restylefloat{table}
\usepackage{booktabs}
\usepackage{enumitem} 
\usepackage{bbold}
\usepackage{amsthm}
\usepackage{slashed}

\usepackage{etoolbox} 
\usepackage{lipsum} 
\usepackage[capitalize]{cleveref}
\usepackage{multirow}
\usepackage[caption=false]{subfig}
\renewcommand\[{\begin{equation}}
\renewcommand\]{\end{equation}}

\newcommand{\ba}{\begin{eqnarray}}
\newcommand{\ea}{\end{eqnarray}}

\makeatletter

\appto{\appendix}{%
\@ifstar{\def\theequation@prefix{A.}}%
{}%
}
\makeatother

\begin{document}

\title{Generalized ghost--free propagators in nonlocal field theories}

\author{Luca Buoninfante}
\affiliation{Department of Physics, Tokyo Institute of Technology, Tokyo 152-8551, Japan}

\author{Gaetano Lambiase}
\affiliation{INFN Sezione di Napoli, Gruppo collegato di Salerno, I-84084 Fisciano (SA), Italy}
\affiliation{Dipartimento di Fisica "E.R. Caianiello", Universit\`a di Salerno, I-84084 Fisciano (SA), Italy}

\author{Yuichi Miyashita}
\affiliation{Department of Physics, Tokyo Institute of Technology, Tokyo 152-8551, Japan}

\author{Wataru Takebe}
\affiliation{Department of Physics, Tokyo Institute of Technology, Tokyo 152-8551, Japan}

\author{Masahide Yamaguchi}
\affiliation{Department of Physics, Tokyo Institute of Technology, Tokyo 152-8551, Japan}


\begin{abstract}
In this paper we present an iterative method to generate an infinite class of new nonlocal field theories whose propagators are ghost--free. We first examine the scalar field case and show that the pole structure of such generalized propagators possesses the standard two derivative pole and in addition can contain complex conjugate poles which, however, do not spoil at least tree level unitarity as the optical theorem is still satisfied. Subsequently, we define analogous propagators for the fermionic sector which is also devoid of unhealthy degrees of freedom. As a third case, we apply the same construction to gravity and define a new set of theories whose graviton propagators around the Minkowski background are ghost--free. Such a wider class also includes nonlocal theories previously studied, and Einstein's general relativity as a peculiar limit. Moreover, we compute the linearized gravitational potential generated by a static point--like source for several gravitational theories belonging to this new class and show that the nonlocal nature of gravity regularizes the singularity at the origin.
\end{abstract}

\maketitle


\section{Introduction}

Einstein's general relativity (GR) has been the most successful theory of gravity so far, indeed its predictions have been tested to very high precision in the infrared (IR) regime, i.e. at large distances and late times \cite{-C.-M.}, though it still needs to be tested at cosmological scales. However, despite its great achievements, there are still unsolved problems suggesting that Einstein's GR can be only seen as an effective field theory of gravitational interaction, which works very well at low energy but breaks down in the ultraviolet (UV) regime, i.e. at short--distances and high energies. In fact, at the classical level the Einstein-Hilbert Lagrangian, $\sqrt{-g}\mathcal{R},$ is plagued by blackhole and cosmological singularities \cite{Hawking}, while at the quantum level it turns out to be perturbatively non-renormalizable \cite{tHooft:1974toh,Goroff:1985th}. 

The most conservative way to extend GR geometrically is to add terms quadratic in the curvatures to the Einstein-Hilbert action, like for example $\mathcal{R}^2$ and $\mathcal{R}_{\mu\nu}\mathcal{R}^{\mu\nu}.$ This kind of action was shown to be power counting renormalizable in Ref.~\cite{-K.-S.}, but still pathological because of the presence of a massive spin-$2$ ghost degree of freedom which causes Hamiltonian instabilities at the classical level, and breaks the unitarity condition of the S--matrix at the quantum level\footnote{See Refs. \cite{Anselmi:2017yux,Anselmi:2018kgz,Anselmi:2017ygm,Anselmi:2018ibi} in which a new quantization prescription is proposed and shown to preserve unitarity: the ghost is converted into a fake degree of freedom ({\it fakeon}), so that the optical theorem can still hold.}. 

The emergence of ghost fields is related to the fact that the field equations contain higher order time derivatives \cite{Ostrogradsky:1850fid}. In the last decades, it was realized that ghost--like degrees of freedom can be still avoided if the derivative order is not finite but {\it infinite}. Indeed, by constructing the quadratic part of the action in terms of  non--polynomial differential operators, like $e^{\Box/\mathcal{M}^2},$ with $\mathcal{M}$ being a new fundamental scale, one can prevent the presence of unhealthy poles in the particle spectrum as initially noticed in \cite{Krasnikov,Kuzmin,Moffat,Tomboulis:1997gg}. The presence of non-polynomial derivatives makes the action {\it nonlocal}, and this kind of nonlocal models were already studied in the early fifties to deal with UV divergences in loop integrals, see Refs. \cite{efimov}. This possibility turned out to be very promising and has motivated a deeper investigation of this unexplored sector of {\it nonlocal (or infinite derivative) field theories}. 

First relevant applications of infinite derivative field theories in a gravitational context were made in Refs. \cite{Biswas:2005qr,Modesto:2011kw,Biswas:2011ar,Biswas:2016etb} in which a stable and unitary quadratic curvature theory of gravity was constructed around maximally symmetric background; see also Ref.~\cite{SravanKumar:2019eqt} for a more general treatment including some non--maximally symmetric spacetime. It was also noticed that the presence of nonlocality can regularize infinities and many efforts have been made towards the resolution of black hole \cite{Modesto:2011kw,Biswas:2011ar,Biswas:2013cha,Edholm:2016hbt,Frolov:2015bia,Frolov,Frolov:2015usa,Buoninfante:2018xiw,Koshelev:2018hpt,Buoninfante:2018rlq,Buoninfante:2018stt,Buoninfante:2018xif,Boos:2018bxf,Kilicarslan:2018yxd,Kilicarslan:2019njc} and cosmological \cite{Biswas:2005qr,Biswas:2010zk,Biswas:2012bp,Koshelev:2012qn,Koshelev:2013lfm} singularities. At the quantum level, the high energy behavior of loop integrals has been investigated in \cite{Modesto:2014lga,Talaganis:2014ida,Koshelev:2017ebj,Ghoshal:2017egr} and properties of causality  and unitarity in \cite{Tomboulis:2015gfa,Buoninfante:2018mre} and \cite{sen-epsilon,carone,Briscese:2018oyx,chin}, respectively. Computations of scattering amplitudes were performed in \cite{Biswas:2014yia,Dona:2015tra,Buoninfante:2018gce}, while a detailed study of spontaneous breaking of symmetry with nonlocal interactions in \cite{Gama:2018cda,Hashi:2018kag}. Applications also appeared in the context of astrophysical compact objects \cite{Buoninfante:2019swn,Buoninfante:2019teo}, cosmological inflation \cite{inflation}, thermal field theory \cite{Biswas:2009nx,Biswas:2010xq,Biswas:2010yx,Boos:2019zml}, dark matter \cite{Ghoshal:2018gpq}, supersymmetry \cite{Giaccari:2016kzy,Kimura:2016irk}, Hawking radiation \cite{Boos:2019vcz}, Galilean theories \cite{Buoninfante:2018lnh}, quantum mechanics \cite{Buoninfante:2017kgj,Buoninfante:2017rbw,Boos:2018kir,Boos:2019fbu}, curved Casimir effect \cite{Buoninfante:2018bkc} and neutrino oscillations \cite{Buoninfante:2019der}.

To better understand our framework, let us briefly review the main aspects of such kind of nonlocal field theories. For simplicity, we consider a Lagrangian for a scalar field of the following type (here we adopt the signature convention $(-+++)$ and the Natural Units $\hbar=1=c$):
\begin{equation}
\mathcal{L}=\frac{1}{2}\phi F(\Box)\phi - V(\phi)\,,
\end{equation}
where the differential operator $F(\Box)$ is required to be an entire function of the flat d'Alembertian $\Box=\eta^{\mu\nu}\partial_{\mu}\partial_{\nu}$, so that it has no poles in the whole complex plane\footnote{Note that we do not consider non--analytic differential operators like $1/\Box$ and ${\rm log}(\Box),$ which are also known and studied in the literature; see for instance Refs.~\cite{Bravinsky,Deser:2007jk,Conroy:2014eja,Belgacem:2017cqo,Woodard:2018gfj,belenchia}}, and physically it represents the kinetic operator, while $V(\phi)$ is some generic potential. From the Weierstrass theorem for the entire functions, it follows that the operator $F(\Box)$ can be written as~\cite{conway}
\begin{equation}
F(\Box)=-e^{\gamma(\Box)}\prod\limits_{i=0}^{N}(-\Box+m_i^2)^{r_i}\,,
\end{equation}
where the function $\gamma(\Box)$ is also entire, the constants $m_i^2$ can be complex numbers and stand for the zeroes of the function $F(\Box),$ while $r_i$ is the multiplicity of each $i$--th zero. The natural number $N$ can be either finite or infinite, and counts the zeroes of the kinetic operator, or in other words, the number of distinct poles of the propagator\footnote{To be more precise we should explicitly show the minus sign coming from the Fourier transform, i.e. we should write either $F(-\Box)\leftrightarrow F(p^2) $ or $F(\Box)\leftrightarrow F(-p^2).$ However, to simplify the notation we adopt the convention $F(\Box)\leftrightarrow F(p^2)$ for any of the functions used in this paper, but of course the minus sign is shown in the explicit expressions on the right hand side.}
\begin{equation}
\Pi(p^2)=\frac{-1}{F(p^2)}=\frac{1}{f(p^2)}\frac{1}{p^2+m^2}\,,\label{momentum space F-f}
\end{equation}
where the function $f(p^2)$ may possess additional zeroes which would contribute to the propagator as extra poles due to higher derivatives.

It is well known that in order to avoid instabilities, i.e. ghost modes, the number of real poles and its multiplicity cannot exceed one. For instance, one possibility in order to avoid Hamiltonian instabilities and preserve unitarity is \cite{Krasnikov,Tomboulis:1997gg,Biswas:2005qr,Biswas:2011ar,Buoninfante:2018mre}
\begin{equation}
F(p^2)=-e^{\gamma(p^2)}(p^2+m^2)\quad \Rightarrow\quad \Pi(p^2)=\frac{e^{-\gamma(p^2)}}{p^2+m^2},\label{exp-propag}
\end{equation}
whose only pole is $p^2=-m^2$ since the exponential of entire function $e^{-\gamma(p^2)}$ does not introduce any extra zeroes in the denominator; we adopt the normalization $e^{\gamma(-m^2)}=1,$ which is usually chosen \cite{Buoninfante:2018mre}. Therefore, although the theory is made up of infinitely higher order derivatives, the number of degrees of freedom and also initial conditions is still finite \cite{Barnaby:2007ve,Calcagni:2018lyd}, that is, two in this case.

Note that similar infinite order differential operators also appear in the context of string field theory \cite{Witten:1985cc,Gross:1987kza,eliezer,Tseytlin:1995uq,Siegel:2003vt} and p-adic string \cite{Freund:1987kt}. 

Very interestingly, in Ref.~\cite{Buoninfante:2018lnh} it was pointed out that tree level unitarity can be preserved even in presence of pairs of complex conjugate poles, indeed it so happens that the optical theorem $2{\rm Im}\left\lbrace T \right\rbrace = T^{\dagger}T$ is still satisfied as the complex conjugate pair does not contribute to the imaginary part of the amplitude $T.$ For instance, a kinetic operator of the type
\begin{equation}
\begin{array}{rl}
F(p^2)=&-e^{\gamma(p^2)}(p^2+m^2)(p^2+iM^2)(p^2-iM^2)\\[3mm]
=&-e^{\gamma(p^2)}(p^2+m^2)(p^4+M^4)\,,
\end{array}
\end{equation}
would give an higher derivative theory which is unitary; see also Ref.~\cite{Anselmi:2017yux,Modesto:2015ozb} for investigations on local field theories with complex conjugate poles known as Lee-Wick theories. 

In Ref.~\cite{Buoninfante:2018lnh} the following nonlocal differential operator was studied:
\begin{eqnarray}
&&\!F(p^2)=\!-\mathcal{M}^2\left(e^{p^2/\mathcal{M}^2}-1\right), \\
&& f(p^2) \equiv - \frac{F(p^2)}{p^2}=\!\mathcal{M}^2\frac{e^{p^2/\mathcal{M}^2}-1}{p^2}\,,
\label{kinetic-inf-compl}
\end{eqnarray}
whose corresponding propagator
\begin{equation}
\Pi(p^2)=\displaystyle\frac{1}{\mathcal{M}^2}\frac{1}{e^{p^2/\mathcal{M}^2}-1} \,,
\label{propag-complex}
\end{equation}
is characterized by an infinite set of pairs of complex conjugate poles: $p^2=i2\pi \mathcal{M}^2\ell,$ with $\ell\in \mathbb{Z}.$ Once can show that the optical theorem is not violated at tree level for the propagator in Eq.~\eqref{propag-complex} \cite{Buoninfante:2018lnh} (see also Subsection \ref{subsec-unitarity} below). The same construction can be also performed in a gravitational context around the Minkowski background, where one can define a ghost--free nonlocal graviton propagator made up of the standard massless pole, $p^2=0,$ plus an infinite set of complex conjugate pairs\footnote{It is worthwhile mentioning that in Refs.~\cite{Abel:2019ufz,Abel:2019zou}, it was noticed that ghost--free nonlocal theories can even admit propagators possessing the standard single real pole plus a brunch cut.}.

Here, our aim is to enlarge the class of ghost--free propagators, first in the scalar and fermionic sectors, and then perform the same construction for gravity, so that we can find a new infinite class of ghost--free nonlocal theories which extends non--trivially others  already known in the literature. We will present an iterative method to generate a infinite tower of nonlocal theories which preserves tree level unitarity despite possessing higher (infinite) order time derivatives. 

The paper is organized as follows. In Section \ref{sec-ghost-free-scalar}, we present an iterative procedure to construct generalized ghost--free propagators in flat spacetime and give a mathematical proof of its validity at any order in the iteration. In Section \ref{sec-fermion}, we perform an analogous construction for the fermionic sector. In Section \ref{sec-quad-gravity}, we briefly review the main aspects of generalized quadratic curvature actions around the Minkowski background. In Section \ref{sec-ghost-free-prop}, we specialize the iterative procedure introduced in Section \ref{sec-ghost-free-scalar} to the gravitational context and construct a new class of generalized ghost--free graviton propagators. Section \ref{sec-nonsing} is devoted to the computation of the gravitational potential generated by a static point--like source for several gravitational theories belonging to this new class. We show that, in the linear regime, the infinite derivative nature of gravity can cure the curvature singularity at the origin. In Section \ref{conclus}, we draw our conclusions.

In the rest of the paper we set $\mathcal{M}=1$ for simplicity.


\section{Ghost--free scalar propagators}\label{sec-ghost-free-scalar}

From now on we only work in the massless case, i.e. we assume that the only real pole of the low--energy theory is $p^2=0,$ as we will be interested in generalizing the same analysis to the gravity sector; however, all our results can be easily adapted to the massive case.

The function $f(p^2)$ in Eq.~\eqref{momentum space F-f} contains all the information on extra poles in the propagator or, in other words, new degrees of freedom. As already mentioned in the Introduction, it is clear that if such a function is an exponential of entire function, 
\begin{equation}
f^{(0)}(p^2)=e^{\gamma(p^2)}\quad \Rightarrow \quad \Pi^{(0)}(p^2)=\frac{e^{-\gamma(p^2)}}{p^2}\,,\label{exp-entire-func}
\end{equation}
then no additional pole appear and the propagator turns out to be ghost--free \cite{Krasnikov,Tomboulis:1997gg,Biswas:2005qr,Biswas:2011ar}. 
From now on, we simply set $\gamma(p^2)=p^2$, that is, $f^{(0)}(p^2)=e^{p^2}$.

Moreover, also the propagator in Eq.~\eqref{propag-complex} was shown to be ghost--free at tree--level \cite{Buoninfante:2018lnh}:
\begin{equation}
\begin{array}{rl}
&\displaystyle f^{(1)}(p^2)=\frac{e^{p^2}-1}{p^2}\quad  \Rightarrow\displaystyle\quad \Pi^{(1)}(p^2)=\frac{1}{e^{p^2}-1}\,,
\end{array}
\label{scalar-propag-complex}
\end{equation}
as no extra real pole appear but only complex conjugate pairs.

We can now notice a very intriguing fact. The propagator in Eq.~\eqref{scalar-propag-complex} is constructed in terms of the function $f^{(0)}(p^2)$ in Eq.~\eqref{exp-entire-func} which defines the propagator $\Pi^{(0)}(p^2),$ indeed the following relation holds true:
\begin{equation}
f^{(1)}(p^2)=\frac{f^{(0)}(p^2)-1}{p^2}\,.\label{iter-0-1}
\end{equation}
Remarkably, this means that we have managed to construct a new ghost--free theory starting from another one which was already known. Thus, it is very natural to ask ourselves whether this property is just a coincidence or it can be generalized and carried on iteratively at higher steps, namely if something like 
\begin{equation}
f^{(n)}(p^2)=c_n\frac{f^{(n-1)}(p^2)-1}{p^2}\,,
\end{equation}
with $c_n$ being constant coefficients to be fixed, can describe new ghost--free theories for any $n\geq 1.$

In the next subsection we will find a positive answer to our question.

\subsection{Iterative procedure and proof for ghost--freeness}

In order to make the notation simpler, let us define $z\equiv p^2.$

First of all, the coefficients $c_n$ can be determined by requiring that the generalized propagator has a well defined low energy limit, i.e. by construction we require
\begin{equation}
\lim\limits_{z\rightarrow0}f^{(n)}(z)=1\quad \Rightarrow\quad c_n=n\,.\label{IR-condition}
\end{equation}
Therefore, the iterative relation reads
\begin{equation}
f^{(0)}(z)=e^z\,,\quad f^{(n)}(z)=n\frac{f^{(n-1)}(z)-1}{z}\,.\label{iterative-relation}
\end{equation}

To avoid  the appearance of unhealthy degrees of freedom, the function $f^{(n)}(z)$ must not possess any real zero, so that not extra real pole is introduced in the propagator besides the massless one $(z=0)\,.$ In fact, we will show that the functions $f^{(n)}(z)$ are always positive for any $z\in \mathbb{R},$ therefore the only extra poles they can possess are complex and have to come in conjugate pairs.

\begin{proof}
	
	In order to prove ghost--freeness, we can recast $f^{(n)}(z)$ in the convenient form
	\begin{equation}
	f^{(n)}(z)=\frac{n!}{z^n}\left(e^z-\sum\limits_{k=0}^{n-1}\frac{z^k}{k!}\right)\,.\label{result-induction}
	\end{equation}
	The validity of Eq.~\eqref{result-induction} can be demonstrated by induction. First, notice that it holds for $n=1$ (see  Eqs.~\eqref{scalar-propag-complex}). As a second part of the induction, we want to show that if it holds for $n,$ then it will also be valid for $n+1.$ Indeed, from  Eq.~\eqref{iterative-relation} we can write:
	\begin{equation}
	\begin{array}{rl}
	f^{(n+1)}(z)=&\displaystyle (n+1)\frac{f^{(n)}(z)-1}{z}    \\[3.5mm]
	=& \displaystyle   \frac{(n+1)!}{z^{n+1}}\left(e^z-\sum\limits_{k=0}^{n-1}\frac{z^k}{k!}-\frac{z^n}{n!}\right)  \\[3.5mm]
	=& \displaystyle \frac{(n+1)!}{z^{n+1}}\left(e^z-\sum\limits_{k=0}^{(n+1)-1}\frac{z^k}{k!}\right)\,,
	\label{induction-proof}
	\end{array}
	\end{equation}
	which proves the validity of Eq.~\eqref{result-induction}.
	
	By making use of the identity
	\begin{equation}
	\sum\limits_{k=0}^{n-1}\frac{z^k}{k!}=e^z\frac{\Gamma(n,z)}{(n-1)!}\,,
	\end{equation}
	with 
	\begin{equation}
	\Gamma(n,z)=\int_{z}^{\infty}dt\,t^{n-1}e^{-t}
	\end{equation}
	being the incomplete gamma function, we can recast Eq.~\eqref{result-induction} in the following elegant form:
	\begin{equation}
	f^{(n)}(z)=e^z z^{-n} g_n(z),\quad g_n(z)\equiv n!-n\,\Gamma(n,z)\,.\label{elegant-form}
	\end{equation}
	To prove that the above function is always positive let us divide the proof in two parts.
	\begin{itemize}
		
		\item $n$ is even: In this case the sign of   $f^{(n)}(z)$ is equal to that of $g_n(z)$ except $z=0$. Moreover, the derivative of $g_n(z)$ reads
		\begin{equation}
		g_n^{\prime}(z)\equiv\frac{d g_n(z)}{dz}=ne^{-z}z^{n-1} \,,\label{derivative-g-odd}
		\end{equation}
		which is positive for $z>0$ and negative for $z<0\,.$ This means that the function $g_n(z)$ has a minimum at $z=0$ where $g_n(0)=0.$ Therefore, $g_n(z)>0$ for any $z\in \mathbb{R}$ except $z=0$,  together with $f^{(n)}(0)=1$, which implies $f^{(n)}(z)>0$ for any $z\in \mathbb{R}$.
		
		\item $n$ is odd: From Eq.~\eqref{derivative-g-odd} it follows that $g_n^{\prime}(z)>0$ for any $z\in \mathbb{R}$ except $z=0$ and since $g_n(0)=0,$ it follows that $g_n(z)>0$ for $z>0,$ while $g_n(z)<0$ for $z<0.$ Therefore, from Eq.~\eqref{elegant-form} and $f^{(n)}(0)=1$, we obtain that $f^{(n)}(z)>0$ for odd $n$ too. 
	\end{itemize}
	
	Hence, we have shown that the functions $f^{(n)}(z)$ are always positive on the real axis so that no extra real pole appear in the propagator, which turns out to be ghost--free at tree level.
	
\end{proof}
In coordinate space, the new class of generalized ghost--free theories is described by the kinetic terms
\begin{equation}
\begin{array}{rl}
\mathcal{L}^{(n)}=&\displaystyle \frac{1}{2}\phi\,f^{(n)}(\Box)\Box \,\phi\,,\\[3mm]
f^{(n)}(\Box)=& \displaystyle \frac{e^{-\Box}}{(-\Box)^{n}}\left[n!-n\,\Gamma\left(n,-\Box\right)\right]\,,
\end{array}\label{n-lagrangian}
\end{equation}
while the generalized propagators in momentum can be expressed in the following compact form:
\begin{equation}
\Pi^{(n)}(p^2)=\displaystyle  \frac{e^{-p^2}p^{2n}}{n!-n\,\Gamma\left(n,p^2\right)}\frac{1}{p^2}\,.
\label{pi^n-scalar}
\end{equation}
Clearly the analogous formulas in the massive case can be obtained by sending $p^2\rightarrow p^2+m^2$ or, equivalently, $\Box\rightarrow \Box-m^2.$


Let us now make some remark on the above result.

\subsubsection{Geometrical interpretation}

We can notice that Eq.~\eqref{iterative-relation} has a precise geometrical meaning: at each $n$--th order the function $f^{(n)}(z)$ is proportional to slope of the function $f^{(n-1)}(z)$ between the points $z$ and $z_0=0$ at which $f^{(n-1)}(z_0)=1,$ and the constant of proportionality is $n.$ Since the starting function is $f^{(0)}(z)=e^z,$ then any $n$--th order is related to the $(n-1)$--th slope of the exponential $e^z:$
\begin{equation}
f^{(n)}(z)=\displaystyle n\frac{(n-1)\frac{(n-2)\frac{3\frac{2\frac{\frac{e^z-1}{z}-1}{z}-1}{\vdots}}{z}-1}{z}-1}{z}\,.\label{geom-meaning}
\end{equation}
In functional analysis there exists the concept of higher order convexity. Any function $h(x)$ is said to be $n$--convex if and only if its $(n-1)$--derivative exists and is convex or, in other words, if its $(n+1)$--derivative exists and is positive. Moreover, one can also show that if a function is $n$--convex, then its $k$--th sloop, with $k=1,\dots,n$ is positive; see Ref.~\cite{rajba-convex} and references therein for details.  

In our case the function $h(x)$ is given by the exponential $e^z$ which is an $n$--convex function for any $n\in \mathbb{N}$ as $d^{(n)}e^z/dz^{n}=e^z>0.$ Therefore, the proof presented above is consistent with already existing mathematical theorems on $n$--convex functions, and represent a specific subcase of a more general topic in functional analysis.

\subsubsection{$n\rightarrow \infty$ limit}

We can also ask whether the class of theories described by the set of functions $\left\lbrace f^{(n)}(z) \right\rbrace $ has a well defined $n\rightarrow \infty$ limit, and which is the corresponding theory, $f^{(\infty)}(z),$ to which it tends. By taking the limit for $n\rightarrow \infty$ we obtain (see Eq.~\eqref{n-lagrangian})
\begin{equation}
f^{(\infty)}(z)\equiv \lim\limits_{n\rightarrow \infty} f^{(n)}(z)=1,
\label{n->infinity}
\end{equation}
which corresponds to the local Klein-Gordon propagator, $\Pi^{(\infty)}(p^2)=1/p^2.$ This means that, if we reinstate the nonlocal energy scale $\mathcal{M},$ there are two possible ways to recover the local limit, either $\mathcal{M}\rightarrow \infty$ or $n\rightarrow \infty.$

This is not so surprising if we think that quantum field theory with the canonical kinetic term of two derivatives is unitary and, therefore, it must consistently belong to the class of ghost--free theories found above.

\subsection{Tree level unitarity}\label{subsec-unitarity}

Given the compact expression for $f^{(n)}(\Box)$ in Eq.~\eqref{n-lagrangian}, we can now discuss the pole structure of the propagator \eqref{pi^n-scalar} in relation to unitarity.

\subsubsection{Optical theorem}

The unitarity condition of the $S$-matrix is defined by the identity 
\begin{equation}
S^{\dagger}S=\mathbb{1},
\label{unitarity-cond}
\end{equation}
which, by introducing the amplitude $T$ such that $S=\mathbb{1}+iT,$ can be also recast in the form
\begin{equation}
i(T^{\dagger}-T)=T^{\dagger}T,
\label{optical theorem}
\end{equation}
known as optical theorem \cite{Anselmi:2016fid}. By introducing $\left| b\right\rangle$ and $\left| a\right\rangle$ as out-- and in--states, respectively, and using the completeness relation, we can write the optical theorem as
\begin{equation}
i\left[\left\langle b|T^{\dagger}|a \right\rangle - \left\langle b|T|a \right\rangle \right]=\sum\limits_n \left\langle b|T^{\dagger}|n \right\rangle\left\langle n|T|a \right\rangle\,.
\label{optical theorem-components}
\end{equation}
Let us now introduce the matrix $M,$ whose components (Feynman diagrams) are defined through the relations
\begin{equation}
\left\langle b|T|a \right\rangle= (2\pi)^4\delta^{(4)}(P_b-P_a)\left\langle b|M|a \right\rangle\,,
\label{feynm-diagr}
\end{equation}
where $P_b$ and $P_a$ are the outgoing and ingoing momenta, respectively. In terms of the Feynman amplitudes, the optical theorem in Eq.\eqref{optical theorem-components} reads
\begin{equation}
\begin{array}{ll}
\displaystyle i\left[\left\langle b|M^{\dagger}|a \right\rangle - \left\langle b|M|a \right\rangle \right]=\displaystyle \sum\limits_n\prod\limits_{l=1}^n\int \frac{d^3k_l}{(2\pi)^3}\frac{1}{2E_l}&\\[3.5mm]
\displaystyle \quad\times (2\pi)^4 \delta^{(4)}\left(P_a-\sum_{l=1}^nk_l\right) \left\langle b|M^{\dagger}|\left\lbrace k_l\right\rbrace  \right\rangle\left\langle \left\lbrace k_l\right\rbrace|M|a \right\rangle\,,&
\end{array}
\label{optical theorem-feynman}
\end{equation}
where we have explicitly written the phase space integral in the completeness relation, i.e.
\begin{equation}
\mathbb{1}=\sum\limits_n\prod\limits_{l=1}^n\int \frac{d^3k_l}{(2\pi)^3}\frac{1}{2E_l} \left|\left\lbrace k_l\right\rbrace  \right\rangle \left\langle \left\lbrace k_l\right\rbrace \right|\,,
\label{complet-integ}
\end{equation}
with the energies $E_l=\sqrt{\vec{p}^2_l+m^2}.$

Note that, in the case $a=b$ (forward scattering amplitude) Eq.\eqref{optical theorem-feynman} reduces to
\begin{equation}
\begin{array}{ll}
\!\displaystyle2{\rm Im}\left\lbrace \left\langle a|M|a \right\rangle  \right\rbrace =\displaystyle \sum\limits_n\prod\limits_{l=1}^n\int \frac{d^3k_l}{(2\pi)^3}\frac{(2\pi)^4}{2E_l}&\\[3.5mm]
\qquad\quad\quad\displaystyle\times\delta^{(4)}\left(P_a-\sum\limits_{l=1}^nk_l\right) \left|\left\langle \left\lbrace k_l \right\rbrace |M|a \right\rangle \right|^2\geq 0 \,,&
\end{array}
\label{forward-scatt}
\end{equation}
which implies that the imaginary part of any forward scattering amplitude has to be non--negative. For example, if we consider a simple amplitude with two constant vertexes and a single internal propagator $\Pi(p^2)$, the optical theorem implies ${\rm Im}\left\lbrace \Pi(p^2) \right\rbrace\geq 0.$ Field theories with higher order time derivatives, like the one in Ref.~\cite{-K.-S.}, are usually characterized by a violation of unitarity as the ghost component of the propagator satisfies the wrong inequality, ${\rm Im}\left\lbrace \Pi_{\rm ghost}(p^2) \right\rbrace <0.$

\subsubsection{Pole structure}

We now want to show explicitly that tree level unitarity is satisfied for the new class of nonlocal theories found above.

First of all, as shown above $p^2=0$ is always the only real pole (or, $p^2=-m^2$ in the massive case). Besides this massless pole, the only other possibility is that pairs of complex conjugate poles appear.

The equation to be satisfied by the additional poles is
\begin{equation}
\Gamma(n,p^2)=(n-1)!\,,\quad n\geq 1;\label{pole-eq}
\end{equation}
while for the propagator with $n=0$ we have no extra poles (see Eq.~\eqref{exp-propag}) \cite{Krasnikov,Biswas:2005qr,Biswas:2011ar}. From Eq.~\eqref{pole-eq}, it follows that if $p^2$ is a solution, then also its complex conjugate $(p^2)^{*}$ will be a solution.  

The case $n=1$ gives $e^{p^2}=1,$ which is the same as Eq.~\eqref{propag-complex} and contains infinite pairs of complex conjugate poles, indeed the propagator can be written as \cite{Buoninfante:2018lnh}
\begin{equation}
\begin{array}{rl}
\Pi^{(1)}(p^2)=&\displaystyle\frac{e^{-\frac{p^2}{2}}}{p^2}\\[3mm]
&\displaystyle+e^{-\frac{p^2}{2}}\sum\limits_{\ell=1}^{\infty}(-1)^{\ell}\!\left(\frac{1}{p^2+i2\pi  \ell}+\frac{1}{p^2-i2\pi \ell}\right)\,.
\end{array}
\label{propag-infinit-pol}
\end{equation}
For other ghost--free theories with generic $n,$ the propagator $\Pi^{(n)}(p^2)$ will exhibit the same feature. For instance, for some $n$ we have checked graphically that infinite pairs of complex conjugate poles also appear. Hence, the most general form of the propagator will be given by
\begin{equation}
\begin{array}{rl}
\Pi^{(n)}(p^2) =& e^{\widetilde{\gamma}(p^2)} \left[ \displaystyle \frac{1}{p^2}+\sum\limits_{\ell=0}^{\infty}\left(\frac{c_\ell^{(n)}}{p^2+(\mu_\ell^{(n)})^2}\right.\right.\\[4.2mm]
&\displaystyle \left.\left.
+\frac{c_\ell^{(n)*}}{p^2+(\mu_\ell^{(n)*})^{2}}\right)\right]\,,
\end{array}
\label{gen-propag-poles}
\end{equation}
where $\widetilde{\gamma}(p^2)$ is an entire function of $p^2$, $(\mu_\ell^{(n)})^{2}$ are the complex poles and $c_\ell^{(n)}$ the residues of the propagator at each pole (except the factor of $e^{\widetilde{\gamma}(p^2)}$). One can easily show that for the propagator in Eq.~\eqref{gen-propag-poles} the optical theorem is still preserved at tree level, i.e. ${\rm Im}\left\lbrace \Pi^{(n)}(p^2) \right\rbrace>0$, by using the fact that (see also Ref.~\cite{Buoninfante:2018lnh})
\begin{equation}
\begin{array}{rl}
&\displaystyle \displaystyle{\rm Im}\left\lbrace \sum\limits_{\ell=0}^{\infty}\left(\frac{c_\ell^{(n)}}{p^2+(\mu_\ell^{(n)})^2}+\frac{c_\ell^{(n)*}}{p^2+(\mu_\ell^{(n)*})^{2}}\right)\right\rbrace =0\,.
\end{array}
\label{ptical theor}
\end{equation}

\subsubsection*{Case $n=1:$ first example}

To better understand the result, let us study in more details the nonlocal model with $n=1$
\begin{equation}
\mathcal{L}^{(1)}=\frac{1}{2}\phi\left(e^{-\Box+m^2}-1\right)\phi\,,
\end{equation}
as in this case we can analytically determine the pole structure; we consider a non--zero mass to be more general. In particular, we investigate the optical theorem for two different interaction terms.

As a first example, we analyze a simple cubic interaction with coupling constant $\lambda,$
\begin{equation}
V(\phi)=\lambda\, \phi^3\,,
\label{cubic-int}
\end{equation}
whose corresponding tree level amplitude for a $2\rightarrow 2$ scattering is given by
\begin{equation}
\left\langle p_3 p_4|M| p_1p_2 \right\rangle =\lambda^2\Pi^{(1)}(p^2)\,,
\label{tree-cubic-int}
\end{equation}
where $p^2=(p_1+p_2)^2=(p_3+p_4)^2$ is the total momentum squared of the two scattered particles, with $p_1,\,p_2$ and $p_3,\,p_4$ being the ingoing and outgoing momenta, respectively.

First, let us consider the case in which in-- and out--states are the same, i.e. let us check the validity of Eq.\eqref{forward-scatt}. The left hand side (LHS) coincides with the imaginary part of the propagator (up to numerical factors):
\begin{equation}
\begin{array}{rl}
{\rm Im}\left\lbrace \left\langle p_1 p_2|M| p_1p_2 \right\rangle\right\rbrace =&\displaystyle \lambda^2{\rm Im}\left\lbrace\frac{e^{-(p^2+m^2)/2}}{p^2+m^2-i\epsilon}\right\rbrace\\[3.6mm]
=&\lambda^2\displaystyle \frac{e^{-(p^2+m^2)/2}\epsilon}{(p^2+m^2)^2+\epsilon^2}\\[3.6mm]
=&\lambda^2\pi \delta (p^2+m^2)>0\,,
\end{array}
\label{imaginary-part-1}
\end{equation}
where to go from first to second line we have used
\begin{equation}
\begin{array}{rl}
\displaystyle {\rm Im}\left\lbrace \sum\limits_{\ell=1}^{\infty}(-1)^{\ell}\left(\frac{1}{p^2+m^2+i2\pi  \ell}\right.\right.&\\[3mm]
\displaystyle \left.\left.+\frac{1}{p^2+m^2-i2\pi \ell}\right)\right\rbrace& =0\,,
\end{array}
\label{zero-imag-pairs}
\end{equation}
while from the second to the third line we have taken the limit $\epsilon \rightarrow 0.$ From Eq.~\eqref{zero-imag-pairs}, we can notice that having complex poles appearing in conjugate pairs is crucial in order to preserve tree level unitarity. Thus, the LHS in Eq.\eqref{forward-scatt} reads
\begin{equation}
{\rm LHS}=2\pi \lambda^2 \delta (p^2+m^2)\,.
\label{LHS-forw}
\end{equation}
As for the right hand side (RHS) of Eq.\eqref{forward-scatt}, we have only one intermediate state (internal line), i.e. $n=1$ and $\left| \left\lbrace k_l\right\rbrace \right\rangle = \left|  k \right\rangle,$ therefore we obtain
\begin{equation}
\begin{array}{rl}
{\rm RHS}= &\displaystyle \int\frac{d^3k}{(2\pi)^3}\frac{(2\pi)^4}{2E_k}\delta^{(4)}(p_1+p_2-k) \left|\left\langle k|M|p_1p_2 \right\rangle \right|^2 \displaystyle \\[3mm]
=&\displaystyle 2\pi \lambda^2 \int d^4k \delta (k^2+m^2)\delta^{(4)}(p_1+p_2-k)  \\[3mm]
= & \displaystyle 2\pi \lambda^2 \delta (p^2+m^2)\,,
\end{array}
\label{RHS-forw}
\end{equation}
which matches with the LHS in Eq.\eqref{LHS-forw}. Note that, to go from the first to the second line of Eq.\eqref{RHS-forw} we have used the identity $\int \frac{d^3k}{(2\pi)^3}\frac{1}{2E_k}=\int \frac{d^4k}{(2\pi)^4} \,2\pi\, \delta (k^2+m^2)$ and $\left\langle k|M|p_1p_2 \right\rangle =\left\langle p_1p_2|M^{\dagger}|k\right\rangle =\lambda .$

So far we have shown the optical theorem only in the case of forward scattering amplitude in Eq.\eqref{forward-scatt}, i.e. with $a=b,$ but by following similar steps we can show the validity of Eq.\eqref{optical theorem-feynman} with $a\neq b$ too. Indeed, since the vertex is a constant we can easily understand that we obtain again 
\begin{equation}
\begin{array}{rl}	
{\rm LHS}=&\displaystyle i\left[\left\langle p_3 p_4|M^{\dagger}| p_1p_2 \right\rangle-\left\langle p_3 p_4|M| p_1p_2 \right\rangle\right]\\[3mm]
=&2\lambda^2{\rm Im}\left\lbrace \Pi^{(1)}(p^2)\right\rbrace\\[3mm]
=&2\pi\lambda^2\delta (p^2+m^2)\,.
\end{array}
\end{equation}
Furthermore, for the right hand side the same happens, indeed we have $\left\langle k|M|p_1p_2 \right\rangle =\left\langle p_3p_4|M^{\dagger}|k\right\rangle =\lambda,$ which still implies ${\rm RHS}=2\pi\lambda^2\delta (p^2+m^2)$. Thus, the optical theorem is satisfied at tree level.

\subsubsection{Case $n=1:$ second example}

As a second example of interaction, we can consider a nonlocal extension of the Galilean scalar Lagrangian \cite{Nicolis:2008in} introduced in \cite{Buoninfante:2018lnh}:
\begin{equation}
V(\phi)= \lambda \left(e^{\Box}-1\right)\phi\frac{\left(e^{\Box}-1\right)}{\Box}\partial_{\mu}\phi\frac{\left(e^{\Box}-1\right)}{\Box}\partial^{\mu}\phi  \,.
\label{cubic-int-galilean}
\end{equation}
The corresponding $2\rightarrow 2$ scattering tree level amplitude now is more involved:
\begin{equation}
\begin{array}{rl}
\left\langle p_3 p_4|M| p_1p_2 \right\rangle =&\displaystyle V(p_1,p_2,p)\Pi^{(1)}(p^2)V(p,p_3,p_4)\,,
\end{array}
\label{tree-cubic-galileon}
\end{equation}
where the vertex reads
\begin{equation}
\begin{array}{rl}
V(p_1,p_2,p)=& \displaystyle \lambda \left(e^{-p_1^2}-1\right)\left(e^{-p_2^2}-1\right)\left(e^{-p^2}-1\right)\\[3mm]
& \displaystyle  \times \left(\frac{p_1\cdot p_2}{p_1^2 p_2^2}+\frac{p_1\cdot p}{p_1^2 p^2}+\frac{p_2\cdot p}{p_2^2 p^2}\right)\,.
\end{array}\label{3vertex}
\end{equation}
Let us prove the optical theorem directly for any $a$ and $b,$ either equal or different. The LHS in Eq.\eqref{optical theorem-feynman} reads 
\begin{equation}
\!\!\begin{array}{rl}	
{\rm LHS}=&\displaystyle i\left[\left\langle p_3 p_4|M^{\dagger}| p_1p_2 \right\rangle-\left\langle p_3 p_4|M| p_1p_2 \right\rangle\right]\\[3mm]
=&\displaystyle 2\pi\lambda^2\!\left(e^{m^2}-1\right)^6\!\! \left(\frac{p_1\cdot p_2}{m^4}-\frac{1}{m^2}\right)^2\delta (p^2+m^2)\,,
\end{array}
\end{equation}
where we have used $p_i^2=-m^2$ for the on--shell momenta and $p_1\cdot p_2=p_3\cdot p_4.$ Moreover, after computing the quantities
\begin{equation}
\begin{array}{rl}	
\left\langle p_3 p_4|M^{\dagger}| k \right\rangle=&\lambda \left(e^{m^2}-1\right)^2\left(e^{-k^2}-1\right) \\[3mm]
&\displaystyle \times \left(\frac{p_3\cdot p_4}{m^4}+\frac{p_3\cdot k}{p_3^2k^2}+\frac{p_4\cdot k}{p_4^2k^2}\right)\,,
\end{array}
\end{equation}
and
\begin{equation}
\begin{array}{rl}	
\left\langle k|M| p_1p_2 \right\rangle=&\lambda \left(e^{m^2}-1\right)^2\left(e^{-k^2}-1\right) \\[3mm]
&\displaystyle \times \left(\frac{p_1\cdot p_2}{m^4}+\frac{p_1\cdot k}{p_1^2k^2}+\frac{p_2\cdot k}{p_2^2k^2}\right)\,,
\end{array}
\end{equation}
we can perform the integral on the right hand side of Eq.\eqref{optical theorem-feynman} and obtain ${\rm LHS}={\rm RHS}$, so that the optical theorem is satisfied.
 
\vspace{0.4cm}

Finally, let us point out that the class of ghost--free theories we have constructed is even wider, indeed we can include generic entire functions $\gamma(\Box)$ in the exponentials in Eqs.(\ref{n-lagrangian},\ref{pi^n-scalar}) by only requiring that $\gamma(0)=0.$ Although no extra real pole appear, for generic entire function the $n\rightarrow \infty$ limit changes.

In this paper we only consider $\gamma(\Box)=-\Box.$


\section{Ghost--free fermion propagators}\label{sec-fermion}

In this Section we define analogous generalized nonlocal Lagrangians in the fermionic sector and also in this case find a new class of propagators which satisfy tree level unitarity. 

It is not difficult to understand that the fermion Lagrangian corresponding to the scalar one in Eq.~\eqref{n-lagrangian} is given by
\begin{equation}
\begin{array}{rl}
\mathcal{L}_F^{(n)}=\displaystyle - \bar{\psi}\,f^{(n)}(\Box)i\gamma^{\mu}\partial_{\mu} \,\psi\,,
\end{array}\label{n-lagrangian-ferm}
\end{equation}
where $\gamma^{\mu}$ are the Dirac matrices. Indeed, the propagator can be written in terms of the scalar one in Eq.~\eqref{pi^n-scalar} as follows:
\begin{equation}
\begin{array}{rl}
\Pi_F^{(n)}(p^2)=&\displaystyle -\frac{1}{f^{(n)}(p^2)}\frac{1}{\gamma^{\mu}p_\mu}\\[3mm]
=&\displaystyle \slashed{p}\, \Pi^{(n)}(p^2)\,,
\end{array}
\label{pi^n-fermion}
\end{equation}
where in the last step we have used the relation $\left\lbrace \gamma^{\mu},\gamma^{\nu}\right\rbrace=-2\eta^{\mu\nu}$ and introduced the notation $\slashed{p}=\gamma^{\mu}p_{\mu}.$ Also in this case, analogous formulas in the massive case can be obtained by sending $\slashed{p}\rightarrow \slashed{p}-m$ or, equivalently, $i\slashed{\partial}\rightarrow i\slashed{\partial}+m.$ From Eq.~\eqref{n->infinity}, it is clear that in the $n\rightarrow\infty$ limit we obtain the Dirac Lagrangian
\begin{equation}
\mathcal{L}_F^{(\infty)}=-\bar{\psi}i\gamma^{\mu}\partial_{\mu}\psi\,.
\label{dirac}
\end{equation}
It is worthwhile mentioning that the fermion Lagrangian in Eq.~\eqref{n-lagrangian-ferm} can be rigorously obtained in a supersymmetric description, indeed it turns out to be the counterpart of the scalar Lagrangian in Eq.~\eqref{n-lagrangian}. Indeed, in Ref.~\cite{Kimura:2016irk} it was shown that for any scalar kinetic operator $f(\Box)\Box,$ the corresponding one for fermions is given by $-f(\Box)i\gamma^{\mu}\partial_{\mu}.$

Note that, the fermion Lagrangian and propagator with $n=0$ has been already considered in the literature \cite{Biswas:2014yia,Ghoshal:2017egr}, while all other theories ($n\geq 1$), to the best of our knowledge, have not been investigated so far.\footnote{In the appendix of Ref.~\cite{Buoninfante:2018lnh}, another nonlocal extension of a Dirac action is proposed. But, it has a massless pole as well as an infinite number of real tachyonic poles, which might lead to tachyonic instabilities.}

It is clear that the pole structure of the propagator in Eq.~\eqref{pi^n-fermion} is the same as \rm{that} in the scalar case, indeed we have a pole at $p^2=0$ (or $p^2=-m^2$ in the massive case), and in addition pairs of complex conjugate poles. As a consequence, this new class of nonlocal fermion Lagrangians possess a propagator which still satisfies tree level unitarity despite the presence of higher time derivatives. 

One can show that all the arguments presented in the previous Section apply to the fermionic sector too. For instance, one can consider the nonlocal kinetic operator in Eq.~\eqref{n-lagrangian-ferm} in presence of a Yukawa interaction term
\begin{equation}
V_F(\phi,\bar{\psi},\psi)=\lambda\, \phi\bar{\psi}\psi\,,
\label{yukawa}
\end{equation}
and show that, in both cases of fermion and scalar internal propagators, the imaginary part of the tree level scattering amplitude is non--negative. Indeed, the crucial role is still played by the presence of complex conjugate pairs which cancel each other when evaluating the imaginary part of the amplitude.


\section{Quadratic curvature actions}\label{sec-quad-gravity}

The iterative procedure introduced above can be also used to generate new ghost--free gravitational theories, as we will show below. In order to set up our framework, in this Section we review the main aspects of quadratic curvature gravity around flat spacetime. 

One can show that the most general parity-invariant and torsion-free action up to linear perturbations around Minkowski background is given by~\cite{Biswas:2011ar}\footnote{Since we are interested in second order metric perturbations of the gravitational action around Minkowski, we can always neglect the term $\mathcal{R}_{\mu\nu\rho\sigma}\mathcal{F}_3(\Box)\mathcal{R}^{\mu\nu\rho\sigma}$ up to this order. Indeed, the following identity is valid for any power $n$ of $\Box:$
	\begin{equation}
	\mathcal{R}_{\mu\nu\rho\sigma}\Box^n\mathcal{R}^{\mu\nu\rho\sigma}=4\mathcal{R}_{\mu\nu}\Box^n\mathcal{R}^{\mu\nu}-\mathcal{R}\Box^n\mathcal{R}+\mathcal{O}(\mathcal{R}^3)+{\rm div},\nonumber
	\end{equation}
	where $\mathcal{O}(\mathcal{R}^3)$ includes higher order contributions $\mathcal{O}(h^3)$ and {\rm div} stands for boundary terms.}:
\begin{equation}
\!S=\displaystyle \!\frac{1}{2\kappa^2}\!\!\int \!\!d^4x\sqrt{-g}\left\lbrace \mathcal{R}\!+\!\frac{1}{2}\left[\mathcal{R}\mathcal{F}_1(\Box)\mathcal{R}+\mathcal{R}_{\mu\nu}\mathcal{F}_2(\Box)\mathcal{R}^{\mu\nu}\right]\right\rbrace 
\label{quad-action}
\end{equation}
where $\kappa:=\sqrt{8\pi G}=1/M_p,$ with $G$ being the Newton constant and $M_p$ the Planck mass, and the differential operators $\mathcal{F}_i(\Box)$ are uniquely determined around Minkowski once the graviton propagator is known~\cite{Modesto:2011kw,Biswas:2011ar} (see also Ref.~\cite{delaCruz-Dombriz:2018aal} for a more general action including torsion). In particular, by perturbing the metric around flat spacetime,
\begin{equation}
g_{\mu\nu}=\eta_{\mu\nu}+\kappa h_{\mu\nu}\,,\label{lin-metric}
\end{equation}
with $h_{\mu\nu}$ being the metric perturbation, we obtain the action up to order $\mathcal{O}(h_{\mu\nu}^2)$ \cite{Biswas:2011ar}:
\begin{equation}
\begin{array}{rl}
S^{(2)}=&\displaystyle \frac{1}{4}\int d^4x\left\lbrace \frac{1}{2}h_{\mu\nu}f(\Box)\Box h^{\mu\nu}-h_{\mu}^{\sigma}f(\Box)\partial_{\sigma}\partial_{\nu}h^{\mu\nu}\right.\\[3mm]
&\displaystyle \,\,\,\,\,\,\,\,\,\,-\frac{1}{2}hg(\Box)\Box h+hg(\Box)\partial_{\mu}\partial_{\nu}h^{\mu\nu}\\[3mm]
&\,\,\,\,\,\,\,\,\,\,\displaystyle\left.+\frac{1}{2}h^{\lambda\sigma}\frac{f(\Box)-g(\Box)}{\Box}\partial_{\lambda}\partial_{\sigma}\partial_{\mu}\partial_{\nu}h^{\mu\nu}\right\rbrace\,;
\label{lin-quad-action}
\end{array}
\end{equation}
$h\equiv\eta_{\mu\nu}h^{\mu\nu}$ is the trace and $\Box=\eta^{\mu\nu}\partial_{\mu}\partial_{\nu}$ the flat d'Alembertian, while
\begin{equation}
f(\Box)=\displaystyle  1+\frac{1}{2}\mathcal{F}_2(\Box)\Box,\quad
g(\Box)= 1-2\mathcal{F}_1(\Box)\Box-\frac{1}{2}\mathcal{F}_2(\Box)\Box\,.\label{relat-form-fact}
\end{equation}
We can obtain the graviton propagator around Minkowski by adding a gauge fixing term and inverting the graviton kinetic operator. One can show that its saturated part reads \cite{Krasnikov,Tomboulis:1997gg,Modesto:2011kw,Biswas:2011ar,Biswas:2013kla}:
\begin{equation}
\Pi_{\mu\nu\rho\sigma}(p^2)=\frac{\mathcal{P}_{\mu\nu\rho\sigma}^2}{f(p^2)p^2}+\frac{\mathcal{P}_{s,\mu\nu\rho\sigma}^0}{(f(p^2)-3g(p^2))p^2},
\end{equation}
where the spin projection operators $\mathcal{P}_{\mu\nu\rho\sigma}^2$ and $\mathcal{P}^0_{s, \mu\nu\rho\sigma}$ \cite{VanNieuwenhuizen:1973fi} project along the spin--$2$ and spin--$0$ components of any symmetric two--rank tensor, respectively. For $f=1=g$ we recover the Einstein-Hilbert propagator,
\begin{equation}
\Pi_{\rm GR, \mu\nu\rho\sigma}(p^2)=\frac{\mathcal{P}_{\mu\nu\rho\sigma}^2}{p^2}-\frac{\mathcal{P}^0_{s, \mu\nu\rho\sigma}}{2p^2}\,,\label{GR-propag}
\end{equation}
which possesses both spin--$2$ and spin--$0$ components off--shell, while on--shell only the spin--$2$ (with $\pm 2$ helicties) survives.

\subsection{Spin-2 graviton propagator}

For simplicity, in what follows we consider gravitational theories whose graviton propagator contains only a spin--$2$ component on--shell. To achieve this we need to demand the necessary condition
\begin{equation}
\displaystyle\mathcal{F}_1(\Box)=-\frac{1}{2}\mathcal{F}_2(\Box)\quad \Leftrightarrow\quad f(\Box)=g(\Box),
\label{form-factors}
\end{equation}
which gives the gravitational action
\begin{equation}
S=\displaystyle \frac{1}{2\kappa^2}\int d^4x\sqrt{-g}\left\lbrace \mathcal{R}-G_{\mu\nu}\mathcal{F}(\Box)\mathcal{R}^{\mu\nu}\right\rbrace, 
\label{quad-action-spin-2}
\end{equation}
with $G_{\mu\nu}=\mathcal{R}_{\mu\nu}-1/2g_{\mu\nu}\mathcal{R}$ being the Einstein tensor and we have redefined 
\begin{equation}
\mathcal{F}(\Box)\equiv\mathcal{F}_1(\Box)=-\frac{f(\Box)-1}{\Box}\,. \label{spin-2-form-factor}
\end{equation}
The corresponding graviton propagator reads
\begin{equation}
\Pi_{\mu\nu\rho\sigma}(p^2)=\frac{1}{f(p^2)}\left(\frac{\mathcal{P}_{\mu\nu\rho\sigma}^2}{p^2}-\frac{\mathcal{P}^0_{s, \mu\nu\rho\sigma}}{2p^2}\right)\,.
\label{propag-spin-2}
\end{equation}
Note that the last expression for the graviton propagator is analogue to the one in Eq.~\eqref{momentum space F-f} for a scalar field, and the function $f(p^2)$ has exactly the same meaning.


\section{Ghost--free graviton propagators}\label{sec-ghost-free-prop}

The iterative procedure introduced in Section \ref{sec-ghost-free-scalar} can be applied straightforwardly to the gravitational context, but in this case the construction is even richer as we can find a relation between the functions $f^{(n)}(p^2)$ and the form factors $\mathcal{F}^{(n-1)}(p^2)$ at each $n$--th order of the iteration.

At $0$--th order, the graviton propagator is given by
\begin{equation}
\begin{array}{rl}
&\displaystyle f^{(0)}(p^2)=e^{p^2}\\[3mm]
\Rightarrow& \Pi^{(0)}_{\mu\nu\rho\sigma}(p^2)=\displaystyle  \frac{e^{-p^2}}{p^2}\left(\mathcal{P}_{\mu\nu\rho\sigma}^2-\frac{1}{2}\mathcal{P}^0_{s, \mu\nu\rho\sigma}\right)\,,
\end{array}
\label{grav-propag-0th}
\end{equation}
and using Eq.~\eqref{spin-2-form-factor}, we also obtain the corresponding form factor
\begin{equation}
\mathcal{F}^{(0)}(p^2)=\frac{e^{p^2}-1}{p^2}\,.
\label{form factor 0th}
\end{equation}
At $1$--st order, we have 
\begin{equation}
\begin{array}{rl}
&\displaystyle f^{(1)}(p^2)=\frac{e^{p^2}-1}{p^2}\\[3mm] \Rightarrow&\displaystyle \Pi^{(1)}_{\mu\nu\rho\sigma}(p^2)=\frac{1}{e^{p^2}-1}\left(\mathcal{P}_{\mu\nu\rho\sigma}^2-\frac{1}{2}\mathcal{P}^0_{s, \mu\nu\rho\sigma}\right)\,,
\end{array}
\label{grav-propag-complex}
\end{equation}
while the form factor reads
\begin{equation}
\begin{array}{rl}
\mathcal{F}^{(1)}(p^2)=&\displaystyle \frac{\frac{e^{p^2}-1}{p^2}-1}{p^2}=\displaystyle \frac{e^{p^2}-1-p^2}{p^4}\,.
\end{array}
\label{form-factor-complex}
\end{equation}
We can now notice that the function $f^{(1)}(p^2)$ coincides with the form factor of the previous iterative order, i.e. $f^{(1)}(p^2)=\mathcal{F}^{(0)}(p^2).$ This means that we have managed to construct a new ghost--free gravitational theory starting from another one which was already known. From the iterative relation in Eq.~\eqref{iterative-relation} obtained in the scalar field case, it follows that for gravity the following relations hold true:
\begin{equation}
\begin{array}{rl}
f^{(0)}(z)=&e^{p^2}\,,\\[3mm]
f^{(1)}(p^2)=&\mathcal{F}^{(0)}(p^2)\,,\\[3mm]
f^{(2)}(p^2)=&2 \mathcal{F}^{(1)}(p^2)\,,\\[3mm]
\vdots&\\[3mm]
f^{(n)}(p^2)=&n\mathcal{F}^{(n-1)}(p^2)\,,\\[3mm]
\vdots&\,
\end{array}
\end{equation}
and at any $n$-th order we obtain a new class of gravitational theories whose propagators
\begin{equation}
\Pi^{(n)}_{\mu\nu\rho\sigma}(p^2)=\frac{1}{f^{(n)}(p^2)}\left(\frac{\mathcal{P}_{\mu\nu\rho\sigma}^2}{p^2}-\frac{\mathcal{P}^0_{s, \mu\nu\rho\sigma}}{2p^2}\right)\,,
\label{propag-spin-2-order-n}
\end{equation}
are ghost--free as no extra real pole appears besides $p^2=0;$ recall that the expression for $f^{(n)}(p^2)$ is given in Eq.\eqref{n-lagrangian}.


\section{Nonsingular gravitational potentials}\label{sec-nonsing}

In this Section we compute the linearized gravitational potential generated by a static point--like source for several theories belonging to the new class constructed above, and show that the presence of non--polynomial operators in the action, i.e. of infinite order derivatives, is crucial in order to regularize the singularity at the origin.

First of all, we can write the following generalized quadratic action whose graviton propagator is ghost--free around the Minkowski background:
\begin{equation}
\begin{array}{rl}
S=&\displaystyle \frac{1}{2\kappa^2}\int d^4x\sqrt{-g}\left\lbrace \mathcal{R}-G_{\mu\nu}\frac{1}{\Box}\mathcal{R}^{\mu\nu}\right.\\[3mm]
&\displaystyle\left.-G_{\mu\nu}\frac{e^{-\Box}}{(-\Box)^{n+1}}\left[n!-n\,\Gamma\left(n,-\Box\right)\right]\mathcal{R}^{\mu\nu}\right\rbrace\,.
\end{array}
\label{quad-action-generalized}
\end{equation}
By varying the corresponding linearized action, i.e. Eq.~\eqref{lin-quad-action} with $f(\Box)=g(\Box)\rightarrow f^{(n)}(\Box),$ up to order $\mathcal{O}(h^2)$ and introducing the interaction with matter through the stress--energy tensor $T_{\mu\nu},$ we obtain the field equations
\begin{equation}
\begin{array}{ll}
\displaystyle f^{(n)}(\Box)\left(\Box h_{\mu\nu}-\partial_{\sigma}\partial_{\nu}h_{\mu}^{\sigma}-\partial_{\sigma}\partial_{\mu}h_{\nu}^{\sigma}\right.&\\[3mm]
\displaystyle \,\,\,\,\,\,\,\,\,\,\,\,\,\,\,\,\,\,\,\,\,\,\left.+\eta_{\mu\nu}\partial_{\rho}\partial_{\sigma}h^{\rho\sigma}+\partial_{\mu}\partial_{\nu}h-\eta_{\mu\nu}\Box h\right)=-2\kappa T_{\mu\nu}\,&
\end{array}
\label{lin-field-eq}
\end{equation}
We choose the Newtonian gauge and express the metric in isotropic coordinates, 
\begin{equation}
ds^2=-(1+2\Phi)dt^2+(1-2\Phi)(dr^2+r^2d\Omega^2),\label{isotr-metric}
\end{equation}
where we have introduced the gravitational potential $\Phi,$ so that  $\kappa h_{00}=-2\Phi<1$, $\kappa h_{ij}=-2\Phi\delta_{ij}<1$ and $\kappa h=-4\Phi.$  In the case of a static point--like source we have $T_{\mu\nu}=m\delta_{\mu}^0\delta_{\nu}^{0}\delta^{(3)}(\vec{r}),$ where $m$ is the mass of the object. Moreover, by imposing the conditions of staticity and spherical symmetry, one can show that Eq.~\eqref{lin-field-eq} reduces to one single modified Poisson equation:
\begin{equation}
f^{(n)}(\nabla^2)\nabla^2\Phi^{(n)}(r)=4\pi Gm\delta^{(3)}(\vec{r}),
\label{field-eq-pot}
\end{equation}
where we have used $\Box\simeq \nabla^2,$ with $\nabla^2$ being the spatial Laplacian.

\begin{figure}[t]
	\includegraphics[scale=0.41]{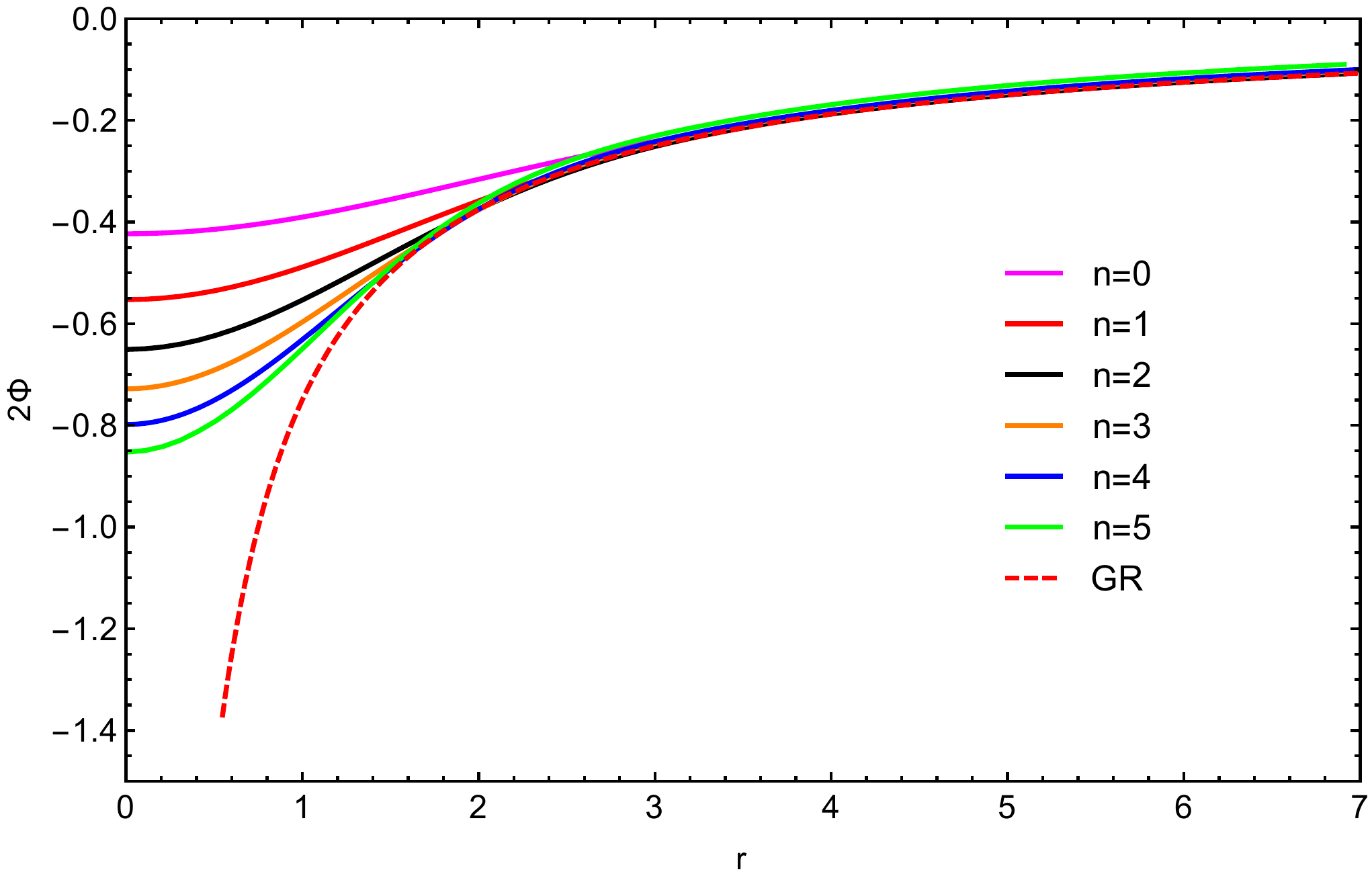}
	\centering
	\protect\caption{We have showed the behavior of the gravitational potential in Eq.~\eqref{fourier-pot} in the cases $n=0$ (magenta line), $n=1$ (red line), $n=2$ (black line), $n=3$ (orange line), $n=4$ (blue line) and $n=5$ (green line), in comparison with the Newtonian potential (red dashed line). For $n=0$ the integral can be computed analytically, while for $n>0$ we have evaluated the integral numerically. For convenience we have set $\mathcal{M}=1$ and $Gm=3/8$.}\label{fig1}
\end{figure}

Eq.~\eqref{field-eq-pot} is highly nonlocal as it contains infinite order derivatives through the function $f^{(n)}(\nabla^2).$ However, we can write its solutions in an integral form by using the Fourier transform method:
\begin{equation}
\begin{array}{rl}
\Phi^{(n)}(r)=&\displaystyle -4\pi Gm\int \frac{d^3k}{(2\pi)^3}\frac{1}{f^{(n)}(k^2)}\frac{e^{i\vec{k}\cdot\vec{r}}}{k^2}\\[3mm]
=&\displaystyle -\frac{2 Gm}{\pi}\frac{1}{r}\int_0^{\infty}dk\frac{1}{f^{(n)}(k^2)}\frac{{\rm sin}(kr)}{k}\\[3mm]
=&\displaystyle -\frac{2 Gm}{\pi}\frac{1}{r}\int_0^{\infty}dk\frac{e^{-k^2}k^{2n-1}}{n!-n\Gamma(n,k^2)}{\rm sin}(kr)
\end{array}
\label{fourier-pot}
\end{equation}
where $k\equiv |\vec{k}|$ and we have used polar coordinates to go from the first to the second line. The integral in Eq.~\eqref{fourier-pot} can be solved analytically only for $n=0,$ which gives $\Phi^{(0)}(r)=-\frac{Gm}{r}{\rm Erf}(\frac{r}{2}),$ while for any other theory with $n\geq1$ we have to do it numerically. The case $n=0$ has been already intensively studied in the literature \cite{Tseytlin:1995uq,Siegel:2003vt,Biswas:2011ar,Buoninfante:2018xiw}, while the theory with $n=1$ has been investigated only recently in Ref.~\cite{Buoninfante:2018lnh}. Instead, to the best of our knowledge, the theories with $n\geq 2$ have been never considered so far, indeed they are totally new and, therefore, worth to analyze.

In Fig.~\ref{fig1}, we have plotted the generalized gravitational potential in Eq.~\eqref{fourier-pot} for several theories, $n=0,1,2,3,4,5\,,$ in comparison with the Newton potential. As it is clear from the plot, we find that for all the examined nonlocal theories the potential is devoid of any singularity, contrarily to the Newtonian one.

We also notice that by increasing the number $n,$ the behavior of the nonlocal potentials approaches the Newtonian one, meaning that theories with higher $n$ describe a stronger gravitational interaction. Note that, such a feature is consistent with the fact that in the $n\rightarrow \infty$ limit we get Einstein's GR; see Eq.~\eqref{n->infinity}. Therefore, we have
\begin{equation}
\Phi^{(\infty)}(r)\equiv \lim\limits_{n\rightarrow \infty} \Phi^{(n)}(r)=-\frac{Gm}{r}\,.
\label{n->infinty-newton}
\end{equation}
Hence, if we reinstate the nonlocal energy scale $\mathcal{M},$ Einstein's GR can be recovered by taking either $\mathcal{M}\rightarrow \infty$ or $n\rightarrow\infty.$

\section{Conclusions}\label{conclus}

In this paper we have defined an iterative procedure to generate an infinite class of ghost-free theories, which non-trivially extends others previously studied and already known in the literature. We have performed the same procedure first for a scalar field, then we applied the same in the fermionic sector and, subsequently, we focused on the gravity. We have classified such new theories in terms of the functions $f^{(n)}(\Box),$ which describes the additional pole structure. All the theories turn out to be nonlocal except that corresponding to $n=\infty$ which coincides with their local limits, namely Klein-Gordon, Dirac and Einstein's GR. We have mathematically proven the ghost-freeness of these generalized propagators by showing that the functions $f^{(n)}(p^2)$ are always positive for any $p^2\in \mathbb{R},$ namely do not have any real zero. The only possible zeroes must be complex and appear in conjugate pairs; however, they do not spoil the optical theorem at tree level.

Moreover, we have computed the generalized gravitational potential generated by a point--like static source and expressed it in a general integral form. In particular, we have shown that in the linear regime the singularity at the origin from which Einstein's GR suffers is now regularized thanks to the smearing feature of nonlocality.

The next would-be task is to find a method to discriminate these new class of theories. In fact, the original motivation of Ref.~\cite{Buoninfante:2018lnh} was to discriminate nonlocal theories from local ones. For this purpose, the local Galilean symmetry was extended to a nonlocal correspondence. We are going to explore this kind of symmetry argument or to derive some consistency relations for the new class of ghost--free theories proposed in this paper, which would be the clue for the discrimination of these theories.

To conclude, let us also mention that future investigations will also focus on the verification of optical theorem at higher order in perturbation theory, by taking into account amplitudes with at least one loop.

\acknowledgements
The authors are very grateful to Teruaki Suyama for enlightening and helpful discussions.  M.~Y. would like to thank Luca Buoninfante and Gaetano Lambiase for their hospitality at University of Salerno, where this work was done. L.~B. acknowledges support from JSPS No. P19324 and KAKENHI Grant-in-Aid for Scientific Research No. JP19F19324. M.~Y.\ is supported in part by JSPS Grant-in-Aid for Scientific Research Numbers 18K18764, MEXT KAKENHI Grant-in-Aid for Scientific Research on Innovative Areas Numbers 15H05888, 18H04579, Mitsubishi Foundation, JSPS and NRF under the Japan-Korea Basic Scientific Cooperation Program, and JSPS Bilateral Open Partnership Joint Research Projects.



\begin{thebibliography}{99}
	
	
			\bibitem{-C.-M.}C. M. Will, Living Rev. Rel. 17, 4
	(2014) {[}arXiv:1403.7377 {[}gr-qc{]}{]}.
	
	\bibitem{Hawking}
	S.W.< Hawking. The Large Scale Structure of Space-Time - 1973. Cambridge University Press. Cambridge, England.
	
	\bibitem{tHooft:1974toh} 
	G.~'t Hooft and M.~J.~G.~Veltman,
	Ann.\ Inst.\ H.\ Poincare Phys.\ Theor.\ A {\bf 20}, 69 (1974).
	
	\bibitem{Goroff:1985th} 
	M.~H.~Goroff and A.~Sagnotti,
	Nucl.\ Phys.\ B {\bf 266}, 709 (1986).
	
	\bibitem{-K.-S.}K. S. Stelle, 
	Phys. Rev. D \textbf{16}, 953 (1977).
	
		\bibitem{Anselmi:2017yux} 
	D.~Anselmi and M.~Piva,
	JHEP {\bf 1706}, 066 (2017),
	[arXiv:1703.04584 [hep-th]].
	D.~Anselmi and M.~Piva,
	Phys.\ Rev.\ D {\bf 96}, no. 4, 045009 (2017),
	[arXiv:1703.05563 [hep-th]].
	
	\bibitem{Anselmi:2018kgz} 
	D.~Anselmi,
	JHEP {\bf 1802}, 141 (2018),
	[arXiv:1801.00915 [hep-th]].
	
	\bibitem{Anselmi:2017ygm} 
	D.~Anselmi,
	JHEP {\bf 1706}, 086 (2017),
	[arXiv:1704.07728 [hep-th]].
	
	
	
	\bibitem{Anselmi:2018ibi} 
	D.~Anselmi and M.~Piva,
	JHEP {\bf 1805}, 027 (2018),
	[arXiv:1803.07777 [hep-th]].
	D.~Anselmi and M.~Piva,
	JHEP {\bf 1811}, 021 (2018),
	[arXiv:1806.03605 [hep-th]].
	D.~Anselmi,
	arXiv:1809.05037 [hep-th].
	
	\bibitem{Ostrogradsky:1850fid} 
	M.~Ostrogradsky,
	Mem.\ Acad.\ St.\ Petersbourg {\bf 6}, no. 4, 385 (1850).
	
	
	\bibitem{Krasnikov}   N. V. Krasnikov, Theor Math. Phys. 73 1184, 1987,
	Teor. Mat. Fiz. 73, 235 (1987).
	
	\bibitem{Kuzmin}Yu. V. Kuzmin, Yad. Fiz. 50, 1630-1635 (1989).
	
	\bibitem{Moffat}
	J. W. Moffat, 
	Phys. Rev. D 41, 1177 (1990). 
	D. Evens, J. W. Moffat, G. Kleppe and R. P. Woodard, 
	Phys. Rev. D 43, 499 (1991).
	
	\bibitem{Tomboulis:1997gg} 
	E.~T.~Tomboulis, 
	hep-th/9702146.
	
	\bibitem{efimov}R.P. Feynman, Phys. Rev. 74, 939 (1948); A. Pais and G. E. Uhlenbeck, Phys. Rev. {\bf 79}, 145 (1950); V. Efimov, Comm. Math. Phys. 5, 42 (1967); V. Efimov, ibid, 7, 138 (1968); V. A. Alebastrov, V. Efimov, Comm. Math. Phys. 31, 1, 1-24 (1973); V. A. Alebastrov, V. Efimov, Comm. Math. Phys. 38, 1, 11-28 (1974); D. A. Kirzhnits (1967) Sov. Phys. Usp. 9 692.
	
	
	\bibitem{Biswas:2005qr} 
	T.~Biswas, A.~Mazumdar and W.~Siegel,
	JCAP {\bf 0603}, 009 (2006)
	[hep-th/0508194].
	
	\bibitem{Modesto:2011kw} 
	L.~Modesto,
	Phys.\ Rev.\ D {\bf 86}, 044005 (2012),
	[arXiv:1107.2403 [hep-th]].
	
	\bibitem{Biswas:2011ar} 
	T.~Biswas, E.~Gerwick, T.~Koivisto and A.~Mazumdar,
	Phys.\ Rev.\ Lett.\  {\bf 108}, 031101 (2012).
	[arXiv:1110.5249 [gr-qc]].
	
	\bibitem{Biswas:2016etb} 
	T.~Biswas, A.~S.~Koshelev and A.~Mazumdar,
	Fundam.\ Theor.\ Phys.\  {\bf 183}, 97 (2016).
	[arXiv:1602.08475 [hep-th]].
	T.~Biswas, A.~S.~Koshelev and A.~Mazumdar,
	Phys.\ Rev.\ D {\bf 95}, no. 4, 043533 (2017).
	[arXiv:1606.01250 [gr-qc]].
	
	
	\bibitem{SravanKumar:2019eqt} 
	K.~Sravan Kumar, S.~Maheshwari and A.~Mazumdar,
	Phys.\ Rev.\ D {\bf 100}, no. 6, 064022 (2019),
	[arXiv:1905.03227 [gr-qc]].
	
	\bibitem{Biswas:2013cha} 
	T.~Biswas, A.~Conroy, A.~S.~Koshelev and A.~Mazumdar,
	Class.\ Quant.\ Grav.\  {\bf 31}, 015022 (2014),
	Erratum: [Class.\ Quant.\ Grav.\  {\bf 31}, 159501 (2014)].
	[arXiv:1308.2319 [hep-th]].
	
	\bibitem{Edholm:2016hbt} 
	J.~Edholm, A.~S.~Koshelev and A.~Mazumdar,
	Phys.\ Rev.\ D {\bf 94}, no. 10, 104033 (2016).
	[arXiv:1604.01989 [gr-qc]].
	V. P. Frolov and A. Zelnikov,  Phys. Rev. D 93, no. 6, 064048 (2016).
	
	
	
	\bibitem{Frolov:2015bia} 
	V.~P.~Frolov, A.~Zelnikov and T.~de Paula Netto,
	JHEP {\bf 1506}, 107 (2015)
	[arXiv:1504.00412 [hep-th]].
	
	\bibitem{Frolov} V.~P.~Frolov,
	Phys.\ Rev.\ Lett.\  {\bf 115}, no. 5, 051102 (2015),
	[arXiv:1505.00492 [hep-th]].
	
	\bibitem{Frolov:2015usa} 
	V.~P.~Frolov and A.~Zelnikov,
	Phys.\ Rev.\ D {\bf 93}, no. 6, 064048 (2016)
	[arXiv:1509.03336 [hep-th]].
	
	
	\bibitem{Buoninfante:2018xiw} 
	L.~Buoninfante, A.~S.~Koshelev, G.~Lambiase and A.~Mazumdar,
	JCAP {\bf 1809} (2018) no.09,  034,
	[arXiv:1802.00399 [gr-qc]].
	
	\bibitem{Koshelev:2018hpt} 
	A.~S.~Koshelev, J.~Marto and A.~Mazumdar,
	Phys.\ Rev.\ D {\bf 98} (2018) no.6,  064023,
	[arXiv:1803.00309 [gr-qc]].
	
	\bibitem{Buoninfante:2018rlq} 
	L.~Buoninfante, A.~S.~Koshelev, G.~Lambiase, J.~Marto and A.~Mazumdar,
	JCAP {\bf 1806} (2018) no.06,  014,
	[arXiv:1804.08195 [gr-qc]].
	
	\bibitem{Buoninfante:2018stt} 
	L.~Buoninfante, G.~Harmsen, S.~Maheshwari and A.~Mazumdar,
	Phys.\ Rev.\ D {\bf 98} (2018) no.8,  084009
	[arXiv:1804.09624 [gr-qc]].
	
	\bibitem{Buoninfante:2018xif}
	L.~Buoninfante, A.~S.~Cornell, G.~Harmsen, A.~S.~Koshelev, G.~Lambiase, J.~Marto and A.~Mazumdar,
	Phys.\ Rev.\ D {\bf 98}, no. 8, 084041 (2018),
	[arXiv:1807.08896 [gr-qc]].
	
	\bibitem{Boos:2018bxf} 
	J.~Boos, V.~P.~Frolov and A.~Zelnikov,
	Phys.\ Rev.\ D {\bf 97}, no. 8, 084021 (2018),
	[arXiv:1802.09573 [gr-qc]].
	
	
	\bibitem{Kilicarslan:2018yxd} 
	E.~Kilicarslan,
	Phys.\ Rev.\ D {\bf 98}, no. 6, 064048 (2018),
	[arXiv:1808.00266 [gr-qc]].
	
	\bibitem{Kilicarslan:2019njc} 
	E.~Kilicarslan,
	Phys.\ Rev.\ D {\bf 99}, no. 12, 124048 (2019),
	[arXiv:1903.04283 [gr-qc]].
	
	
	
	\bibitem{Biswas:2010zk} 
	T.~Biswas, T.~Koivisto and A.~Mazumdar,
	JCAP {\bf 1011}, 008 (2010),
	[arXiv:1005.0590 [hep-th]].
	
	\bibitem{Biswas:2012bp} 
	T.~Biswas, A.~S.~Koshelev, A.~Mazumdar and S.~Y.~Vernov,
	JCAP {\bf 1208}, 024 (2012),
	[arXiv:1206.6374 [astro-ph.CO]].
	
	
	\bibitem{Koshelev:2012qn} 
	A.~S.~Koshelev and S.~Y.~Vernov,
	Phys.\ Part.\ Nucl.\  {\bf 43}, 666 (2012),
	[arXiv:1202.1289 [hep-th]].
	
	\bibitem{Koshelev:2013lfm} 
	A.~S.~Koshelev,
	Class.\ Quant.\ Grav.\  {\bf 30}, 155001 (2013),
	[arXiv:1302.2140 [astro-ph.CO]].
	
	
	\bibitem{Modesto:2014lga} 
	L.~Modesto and L.~Rachwal,
	Nucl.\ Phys.\ B {\bf 889}, 228 (2014),
	[arXiv:1407.8036 [hep-th]].
	L.~Modesto and L.~Rachwał,
	Nucl.\ Phys.\ B {\bf 900}, 147 (2015),
	[arXiv:1503.00261 [hep-th]].
	
	\bibitem{Talaganis:2014ida} 
	S.~Talaganis, T.~Biswas and A.~Mazumdar,
	Class.\ Quant.\ Grav.\  {\bf 32}, no. 21, 215017 (2015),
	[arXiv:1412.3467 [hep-th]].
	
	\bibitem{Ghoshal:2017egr} 
	A.~Ghoshal, A.~Mazumdar, N.~Okada and D.~Villalba,
	Phys.\ Rev.\ D {\bf 97}, no. 7, 076011 (2018)
	[arXiv:1709.09222 [hep-th]].
	
	\bibitem{Koshelev:2017ebj} 
	A.~S.~Koshelev, K.~Sravan Kumar, L.~Modesto and L.~Rachwał,
	Phys.\ Rev.\ D {\bf 98}, no. 4, 046007 (2018),
	[arXiv:1710.07759 [hep-th]].
	
	
	\bibitem{Tomboulis:2015gfa} 
	E.~T.~Tomboulis,
	Phys.\ Rev.\ D {\bf 92}, no. 12, 125037 (2015)
	[arXiv:1507.00981 [hep-th]].
	
	
	\bibitem{Buoninfante:2018mre} 
	L.~Buoninfante, G.~Lambiase and A.~Mazumdar,
	Nucl.\ Phys.\ B {\bf 944}, 114646 (2019),
	[arXiv:1805.03559 [hep-th]].
	
	\bibitem{sen-epsilon}R. Pius and A. Sen, 
	JHEP {\bf 1610}, 024 (2016)
	Erratum: [JHEP {\bf 1809}, 122 (2018)],
	[arXiv:1604.01783 [hep-th]].
	
	
	
	\bibitem{carone}C. D. Carone, 
	Phys. Rev. D 95, 045009 (2017), 
	{[}arXiv:1605.02030v3 [hep-th]{]}.
	
	\bibitem{Briscese:2018oyx} 
	F.~Briscese and L.~Modesto,
	Phys.\ Rev.\ D {\bf 99}, no. 10, 104043 (2019),
	[arXiv:1803.08827 [gr-qc]].
	
	
	
	\bibitem{chin}
	P.~Chin and E.~T.~Tomboulis,
	JHEP {\bf 1806} (2018) 014,
	[arXiv:1803.08899 [hep-th]].
	
	
	
	\bibitem{Biswas:2014yia} 
	T.~Biswas and N.~Okada,
	Nucl.\ Phys.\ B {\bf 898}, 113 (2015),
	[arXiv:1407.3331 [hep-ph]].
	
	\bibitem{Dona:2015tra} 
	P.~Donà, S.~Giaccari, L.~Modesto, L.~Rachwal and Y.~Zhu,
	JHEP {\bf 1508}, 038 (2015),
	[arXiv:1506.04589 [hep-th]].
	
	\bibitem{Buoninfante:2018gce} 
	L.~Buoninfante, A.~Ghoshal, G.~Lambiase and A.~Mazumdar,
	Phys.\ Rev.\ D {\bf 99}, no. 4, 044032 (2019),
	[arXiv:1812.01441 [hep-th]].
	
	
	
	\bibitem{Gama:2018cda} 
	F.~S.~Gama, J.~R.~Nascimento, A.~Y.~Petrov and P.~J.~Porfirio,
	arXiv:1804.04456 [hep-th].
	
	\bibitem{Hashi:2018kag} 
	M.~N.~Hashi, H.~Isono, T.~Noumi, G.~Shiu and P.~Soler,
	JHEP {\bf 1808}, 064 (2018)
	[arXiv:1805.02676 [hep-th]].

	
	\bibitem{Buoninfante:2019swn} 
	L.~Buoninfante and A.~Mazumdar,
	Phys.\ Rev.\ D {\bf 100}, no. 2, 024031 (2019),
	[arXiv:1903.01542 [gr-qc]].
	
	\bibitem{Buoninfante:2019teo} 
	L.~Buoninfante, A.~Mazumdar and J.~Peng,
	Phys.\ Rev.\ D {\bf 100}, no. 10, 104059 (2019),
	arXiv:1906.03624 [gr-qc].
	
	
	
	\bibitem{inflation}
	N. Barnaby, T. Biswas and J. M. Cline, 
	JHEP 0704 (2007) 056, [hep-th/0612230];
	T.~Biswas, R.~Brandenberger, A.~Mazumdar and W.~Siegel,
	JCAP {\bf 0712}, 011 (2007),
	[hep-th/0610274];
	T.~Biswas and A.~Mazumdar,
	Class.\ Quant.\ Grav.\  {\bf 31}, 025019 (2014),
	[arXiv:1304.3648 [hep-th]];
	A.~S.~Koshelev, L.~Modesto, L.~Rachwal and A.~A.~Starobinsky,
	JHEP {\bf 1611}, 067 (2016),
	[arXiv:1604.03127 [hep-th]];
	A.~S.~Koshelev, K.~Sravan Kumar and A.~A.~Starobinsky,
	JHEP {\bf 1803}, 071 (2018),
	[arXiv:1711.08864 [hep-th]];
	K.~Sravan Kumar and L.~Modesto,
	arXiv:1810.02345 [hep-th];
	A.~S.~Koshelev, K.~Sravan Kumar and P.~Vargas Moniz,
	Phys.\ Rev.\ D {\bf 96}, no. 10, 103503 (2017)
	[arXiv:1604.01440 [hep-th]].
	
	\bibitem{Biswas:2009nx} 
	T.~Biswas, J.~A.~R.~Cembranos and J.~I.~Kapusta,
	Phys.\ Rev.\ Lett.\  {\bf 104}, 021601 (2010),
	[arXiv:0910.2274 [hep-th]].
	
	\bibitem{Biswas:2010xq} 
	T.~Biswas, J.~A.~R.~Cembranos and J.~I.~Kapusta,
	JHEP {\bf 1010}, 048 (2010),
	[arXiv:1005.0430 [hep-th]].
	
	\bibitem{Biswas:2010yx} 
	T.~Biswas, J.~A.~R.~Cembranos and J.~I.~Kapusta,
	Phys.\ Rev.\ D {\bf 82}, 085028 (2010),
	[arXiv:1006.4098 [hep-th]].
	
	\bibitem{Boos:2019zml} 
	J.~Boos, V.~P.~Frolov and A.~Zelnikov,
	Phys.\ Lett.\ B {\bf 793}, 290 (2019),
	[arXiv:1904.07917 [hep-th]].
	
	\bibitem{Ghoshal:2018gpq} 
	A.~Ghoshal,
	Int.\ J.\ Mod.\ Phys.\ A {\bf 34}, no. 24, 1950130 (2019),
	[arXiv:1812.02314 [hep-ph]].
	
		
	\bibitem{Giaccari:2016kzy} 
	S.~Giaccari and L.~Modesto,
	Phys.\ Rev.\ D {\bf 96}, no. 6, 066021 (2017),
	[arXiv:1605.03906 [hep-th]].
	
	
	\bibitem{Kimura:2016irk} 
	T.~Kimura, A.~Mazumdar, T.~Noumi and M.~Yamaguchi,
	JHEP {\bf 1610}, 022 (2016),
	[arXiv:1608.01652 [hep-th]].
	
	
	\bibitem{Boos:2019vcz} 
	J.~Boos, V.~P.~Frolov and A.~Zelnikov,
	Phys.\ Rev.\ D {\bf 100}, no. 10, 104008 (2019),
	[arXiv:1909.01494 [hep-th]].
	
	
	\bibitem{Buoninfante:2018lnh} 
	L.~Buoninfante, G.~Lambiase and M.~Yamaguchi,
	Phys.\ Rev.\ D {\bf 100}, no. 2, 026019 (2019),
	[arXiv:1812.10105 [hep-th]].
	
	
	\bibitem{Buoninfante:2017kgj} 
	L.~Buoninfante, G.~Lambiase and A.~Mazumdar,
	Nucl.\ Phys.\ B {\bf 931}, 250 (2018),
	[arXiv:1708.06731 [quant-ph]].
	
	
	
	\bibitem{Buoninfante:2017rbw} 
	L.~Buoninfante, G.~Lambiase and A.~Mazumdar,
	Eur.\ Phys.\ J.\ C {\bf 78}, no. 1, 73 (2018),
	[arXiv:1709.09263 [gr-qc]].
	
	\bibitem{Boos:2018kir} 
	J.~Boos, V.~P.~Frolov and A.~Zelnikov,
	Phys.\ Lett.\ B {\bf 782}, 688 (2018),
	[arXiv:1805.01875 [hep-th]].
	
	\bibitem{Boos:2019fbu} 
	J.~Boos, V.~P.~Frolov and A.~Zelnikov,
	Phys.\ Rev.\ D {\bf 99}, no. 7, 076014 (2019),
	[arXiv:1901.07096 [hep-th]].
	
	
	
	
	\bibitem{Buoninfante:2018bkc} 
	L.~Buoninfante, G.~Lambiase, L.~Petruzziello and A.~Stabile,
	Eur.\ Phys.\ J.\ C {\bf 79}, no. 1, 41 (2019),
	[arXiv:1811.12261 [gr-qc]].
	
	\bibitem{Buoninfante:2019der} 
	L.~Buoninfante, G.~G.~Luciano, L.~Petruzziello and L.~Smaldone,
	Phys.\ Rev.\ D {\bf 101}, no. 2, 024016 (2020),
	[arXiv:1906.03131 [gr-qc]].
	
	
	
		\bibitem{Bravinsky}		
	A. O. Barvinsky and G. A. Vilkovisky, 
	Phys. Rept. 119 (1985) 1-74.
	
	\bibitem{Deser:2007jk} 
	S.~Deser and R.~P.~Woodard,
	Phys.\ Rev.\ Lett.\  {\bf 99}, 111301 (2007).
	
    \bibitem{Conroy:2014eja} 
    A.~Conroy, T.~Koivisto, A.~Mazumdar and A.~Teimouri,
    Class.\ Quant.\ Grav.\  {\bf 32}, no. 1, 015024 (2015),
    [arXiv:1406.4998 [hep-th]].
	
	
	\bibitem{Belgacem:2017cqo} 
	E.~Belgacem, Y.~Dirian, S.~Foffa and M.~Maggiore,
	JCAP {\bf 1803}, no. 03, 002 (2018).
	
	\bibitem{Woodard:2018gfj} 
	R.~P.~Woodard,
	Universe {\bf 4}, no. 8, 88 (2018).
	
	\bibitem{belenchia}A. Belenchia, D. M. T. Benincasa, S. Liberati, 
	JHEP 1503 (2015) 036, {[}arXiv:1411.6513v2 [gr-qc]{]}. 
	A. Belenchia, D. M. T. Benincasa, E. Martin-Martinez, M. Saravani, Phys. Rev. D 94, 061902(R) (2016), {[}arXiv:1605.03973 [quant-ph]{]}.
	
	
	\bibitem{conway}J.~B.~Conway, Functions of One Complex Variable I, 2nd ed., Springer-Verlag New York (1995), DOI:10.1007/978-1-4612-6313-5.
	
	
	
	
			
	\bibitem{Barnaby:2007ve} 
	N.~Barnaby and N.~Kamran,
	JHEP {\bf 0802}, 008 (2008),
	[arXiv:0709.3968 [hep-th]].
	
	
	\bibitem{Calcagni:2018lyd} 
	G.~Calcagni, L.~Modesto and G.~Nardelli,
	JHEP {\bf 1805}, 087 (2018)
	Erratum: [JHEP {\bf 1905}, 095 (2019)],
	[arXiv:1803.00561 [hep-th]].
	
	
	\bibitem{Witten:1985cc} 
	E.~Witten,
	Nucl.\ Phys.\ B {\bf 268}, 253 (1986).
	
	
		
	
	\bibitem{Gross:1987kza} 
	D.~J.~Gross and P.~F.~Mende,
	Phys.\ Lett.\ B {\bf 197}, 129 (1987);
	Nucl.\ Phys.\ B {\bf 303}, 407 (1988).
	
	
	
	\bibitem{eliezer}D.A. Eliezer, R.P. Woodard, Nucl. Phys. B, 325 389 (1989).
	
	\bibitem{Tseytlin:1995uq}
	A.~A.~Tseytlin,
	Phys.\ Lett.\ B {\bf 363} (1995) 223,
	[hep-th/9509050].
	
	\bibitem{Siegel:2003vt} 
	W.~Siegel,
	hep-th/0309093.
	
	\bibitem{Freund:1987kt} 
	P.~G.~O.~Freund and M.~Olson,
	Phys.\ Lett.\ B {\bf 199}, 186 (1987);
	L.~Brekke, P.~G.~O.~Freund, M.~Olson and E.~Witten,
	Nucl.\ Phys.\ B {\bf 302}, 365 (1988);
	P.~G.~O.~Freund and E.~Witten,
	Phys.\ Lett.\ B {\bf 199}, 191 (1987);
	P. H. Frampton and Y. Okada,  
	Phys.Rev., vol. D37, pp. 3077-3079, 1988;
	B.~Dragovich, A.~Y.~Khrennikov, S.~V.~Kozyrev and I.~V.~Volovich,
	Anal.\ Appl.\  {\bf 1}, 1 (2009)
	[arXiv:0904.4205 [math-ph]].
	
	
		\bibitem{Lee:1969fy} 
	T.~D.~Lee and G.~C.~Wick,
	Nucl.\ Phys.\ B {\bf 9}, 209 (1969).
	
	
	\bibitem{Modesto:2015ozb} 
	L.~Modesto and I.~L.~Shapiro,
	Phys.\ Lett.\ B {\bf 755}, 279 (2016), [arXiv:1512.07600 [hep-th]];
	L.~Modesto,
	Nucl.\ Phys.\ B {\bf 909}, 584 (2016),
	[arXiv:1602.02421 [hep-th]];
	B.~L.~Giacchini,
	Phys.\ Lett.\ B {\bf 766}, 306 (2017),
	[arXiv:1609.05432 [hep-th]];
	A.~Accioly, B.~L.~Giacchini and I.~L.~Shapiro,
	Phys.\ Rev.\ D {\bf 96}, no. 10, 104004 (2017),
	[arXiv:1610.05260 [gr-qc]].

	
	\bibitem{Abel:2019ufz} 
	S.~Abel and N.~A.~Dondi,
	JHEP {\bf 1907}, 090 (2019),
	[arXiv:1905.04258 [hep-th]].
	
\bibitem{Abel:2019zou} 
S.~Abel, L.~Buoninfante and A.~Mazumdar,
JHEP {\bf 2001}, 003 (2020),
[arXiv:1911.06697 [hep-th]].
	
	
	\bibitem{rajba-convex}T. Rajba, 
	J. Math. Anal. Appl., {\bf 379} (2) (2011), 736–747.
	
		\bibitem{Anselmi:2016fid} 
	D.~Anselmi,
	Phys.\ Rev.\ D {\bf 94}, 025028 (2016),
	[arXiv:1606.06348 [hep-th]].
	
	
	\bibitem{Nicolis:2008in} 
	A.~Nicolis, R.~Rattazzi and E.~Trincherini,
	Phys.\ Rev.\ D {\bf 79}, 064036 (2009),
	[arXiv:0811.2197 [hep-th]].
	
	
		
	\bibitem{delaCruz-Dombriz:2018aal} 
	Á.~de la Cruz-Dombriz, F.~J.~Maldonado Torralba and A.~Mazumdar,
	Phys.\ Rev.\ D {\bf 99}, no. 10, 104021 (2019),
	[arXiv:1812.04037 [gr-qc]].
	
	\bibitem{Biswas:2013kla} 
	T.~Biswas, T.~Koivisto and A.~Mazumdar,
	arXiv:1302.0532 [gr-qc].
	
	
	\bibitem{VanNieuwenhuizen:1973fi} 
	P.~Van Nieuwenhuizen,
	Nucl.\ Phys.\ B {\bf 60}, 478 (1973).
	

	
	
	
\end{thebibliography}
\end{document}